\documentclass[letterpaper, 10 pt, conference]{ieeeconf} 

\IEEEoverridecommandlockouts               

\usepackage{graphics} 
\usepackage{epsfig} 
\usepackage{times} 
\usepackage{amsmath} 
\usepackage{amssymb} 

\usepackage{tikz}
\usepackage{pgfplots}
\pgfplotsset{compat=1.17}

\usepackage{cite}
\usepackage{bm} 
\usepackage{setspace}
\renewcommand{\baselinestretch}{0.971}

\newtheorem{proposition}{Proposition}

\newtheorem{definition}{Definition}

\newtheorem{corollary}{Corollary}

\usepackage{textcomp}
\usepackage{hyperref}
\usepackage{lipsum}
\newcommand\copyrighttext{
	\footnotesize \textcopyright 2021 IEEE.  Personal use of this material is permitted.  Permission from IEEE must be obtained for all other uses, in any current or future media, including reprinting/republishing this material for advertising or promotional purposes, creating new collective works, for resale or redistribution to servers or lists, or reuse of any copyrighted component of this work in other works.
	DOI: \href{https://doi.org/10.1109/CDC45484.2021.9683158}{10.1109/CDC45484.2021.9683158}}
\newcommand\copyrightnotice{
	\begin{tikzpicture}[remember picture,overlay]
		\node[anchor=south,yshift=10pt] at (current page.south) {\fbox{\parbox{\dimexpr\textwidth-\fboxsep-\fboxrule\relax}{\copyrighttext}}};
	\end{tikzpicture}%
}

\title{\LARGE \bf Constraint Removal for MPC with Performance Preservation\\ and a Hyperthermia Cancer Treatment Case~Study}

\author{S.A.N. Nouwens$^{1}$, B. de Jager$^{1}$, M.M. Paulides$^{2,3}$, W.P.M.H. Heemels$^{1}$
\thanks{This research is supported by KWF Kankerbestrijding and NWO Domain AES, as part of their joint strategic research programme: Technology for Oncology II. The collaboration project is co-funded by the PPP Allowance made available by Health$\sim$Holland, Top Sector Life Sciences \& Health, to stimulate public-private partnerships.}
\thanks{$^{1}$Control Systems Technology, Department of Mechanical Engineering, Eindhoven University of Technology, Eindhoven, The Netherlands
}%
\thanks{$^{2}$Electromagnetics for Care \& Cure, Department of Electrical Engineering, Eindhoven University of Technology, Eindhoven, The Netherlands}%
\thanks{$^{3}$Department of	Radiotherapy, Erasmus University Medical Center Cancer Institute, Rotterdam, The Netherlands}%
}

\begin{document}

\maketitle
\thispagestyle{empty}
\pagestyle{empty}

\copyrightnotice
\begin{abstract}
Model predictive control (MPC) is an optimization-based control strategy with broad industrial adoption. Unfortunately, the required computation time to solve the receding-horizon MPC optimization problem can become prohibitively large for many applications with a large number of state constraints. This large number of state constraints can, for instance, originate from spatially discretizing a partial differential equation of which the solution has to satisfy constraints over the full spatial domain. This is particularly the case in MPC for RF-based hyperthermia cancer treatments, which forms a strong  motivation for this study. To address this problem, we propose a novel constraint-adaptive MPC framework for linear discrete-time systems. In this framework, we select at each time-step a subset of the state constraints that are included in the optimization problem, thereby reducing the online computational burden. Critically, our framework guarantees the same closed-loop performance, recursive feasibility, and constraint satisfaction properties as the original (non-reduced) MPC scheme. We achieve this result by efficiently  exploiting reachable set computations and the MPC cost function. We will demonstrate our novel method using a hyperthermia cancer treatment case study showing a two-orders of magnitude improvement in computation time, with identical closed-loop performance as the original (non-reduced) MPC scheme. 
\end{abstract}

\section{Introduction}
Model predictive control (MPC) is a successful, optimization-based, control technology \cite{Mayne2014}. Generally, the control inputs are obtained by recursively optimizing a finite-horizon cost function subject to input and state constraints and system dynamics. The explicit inclusion of state and inputs constraints and the capability to handle multi-input multi-output systems are distinct advantages of MPC. However, solving an optimization problem at each time-step can prohibit the real-time feasibility of the controller for computationally complex scenarios. Particularly, in applications requiring the control of systems modeled by partial differential equations with constraints on the continuous spatial domain, such as the heat equation for hyperthermia cancer treatments, the real-time application of MPC can be problematic.

Efforts to improve the computational aspects of MPC are commonplace with among others, explicit MPC, model reduction, and tailored numerical solvers, see, e.g., \cite{Bemporad2002,Hovland2006,Genuit2011,Jerez2011,Rawlings2019}, and the references therein. However, our focus in this paper is on systems with large-scale constraints. For such systems, the aforementioned techniques are typically not highly effective. In particular, interior-point methods are reported to scale linearly with the number of inequality constraints \cite{Jerez2011}. As a result, systems with many constraints motivate so-called constraint removal techniques, which are roughly separated into offline \cite{Paulraj2010}, and online methods \cite{Jost2013,Jost2015,Nouwens2021}. Offline methods can be prohibitively complex to compute and possibly do not enable real-time MPC, as constraints can only be removed if they are redundant with respect to all feasible states. In contrast, online techniques can typically remove more constraints as they exploit the knowledge of the current state of the system. However, online methods use a so-called pre-solve step whose \textit{additional} computation time cannot be ignored and should be limited. Although the existing online constraint removal techniques are effective in various scenarios, they still require extensive computational effort for large-scale problems, or do not provide performance guarantees. 

In this paper, we propose a novel online constraint removal method for MPC of linear discrete-time systems that is computationally tractable and guarantees the same closed-loop performance, recursive feasibility, and constraint satisfaction as the original MPC scheme. To this end, we exploit properties inherent to the system by using fast reachability analysis and the optimal control sequence from the previous time step to systematically remove constraints from the optimization problem. We will show the strengths of our new framework using a hyperthermia cancer treatment case study \cite{Deenen2020,Hendrikx2018,VanderZee2002,Hensley2015}, where we obtain a two-orders of magnitude reduction in computational time. Note that existing work on MPC for hyperthermia is largely focused on ultrasound heating and localized 2D models, which limits their application to small tumors \cite{Deenen2020,Hensley2015}. In contrast, radio frequency based (RF) hyperthermia, used for larger tumors, requires large-scale 3D models due to the non-localized heating from the actuators \cite{Hendrikx2018}. The non-local heating also introduces the need for temperature constraints on the whole spatial domain to ensure patient safety. For these reasons, MPC for RF-based hyperthermia is computationally challenging.

The remainder of this paper is structured as follows. We start by introducing the preliminaries in Section~\ref{sec:preliminaries}. In Section~\ref{sec:reduction}, we provide our ca-MPC scheme that exploits reachable sets and the cost function to remove constraints without changing the optimizer of the MPC problem, after which, we provide computationally cheap methods to remove state constraints in Section~\ref{sec:removal}. The hyperthermia cancer treatment case study is presented in Section~\ref{sec:casestudy}. Finally, conclusions are stated in Section~\ref{sec:conclusion}.


\section{Preliminaries}\label{sec:preliminaries}
In this section, we introduce the linear system to be controlled, a typical MPC setup, the forward and backward reachable sets and the cost function level sets. These sets form the basis of our constraint removal framework.
\subsection{System} 
In this paper, we consider plants that are described by the discrete-time linear time-invariant (LTI) system
\begin{align}\label{eq:system_description}
	&\bm{x}_{k+1} = \bm{Ax}_k + \bm{Bu}_k.
\end{align}
Here, $\bm{x}_k\in\mathbb{R}^n$ and $\bm{u}_k\in\mathbb{R}^m$ denote the plant states and the inputs, respectively, at discrete time $k\in\mathbb{N}$. Furthermore, $\bm{A}$ and $\bm{B}$ are real matrices of appropriate dimensions. The system \eqref{eq:system_description} is subject to polyhedral state and input constraints that are given by
\begin{subequations}\label{eq:state_and_input_constraints}
	\begin{align} 
		\bm{x}_k &\in \{\bm{x}\in\mathbb{R}^n\mid \bm{Cx}\leq \bm{b}\} &=: \mathbb{X},\hspace{2cm}\\
		\bm{u}_k &\in \{\bm{u}\in\mathbb{R}^m\mid \bm{C}_u\bm{u} \leq \bm{b}_u\}&=: \mathbb{U}.\hspace{2cm} \label{eq:input_constraints}
	\end{align}
\end{subequations}
Here, $\mathbb{X}$ and $\mathbb{U}$ are assumed to be non-empty polyhedral sets with $\bm{C}\in\mathbb{R}^{n_x\times n}$, $\bm{C}_u\in\mathbb{R}^{n_u\times m}$, $\bm{b}\in\mathbb{R}^{n_x}$, and $\bm{b}_u\in\mathbb{R}^{n_u}$. In this paper we are interested in systems constrained by many state constraints, i.e., $n_x\gg1$.


\subsection{MPC setup} 
Based on the system dynamics \eqref{eq:system_description} and constraints \eqref{eq:state_and_input_constraints}, a common MPC setup, given state $\bm{x}_k$ at time $k\in\mathbb{N}$, can be formulated as 
\begin{subequations}\label{eq:basic_mpc}
	\begin{align}
		\underset{\bm{U}_k}{\text{minimize}}\ \quad&J(\bm{x}_k,\bm{U}_k),\label{eq:basic_mpc_a}\\
		\text{subject to }\quad&\bm{X}_k = \bm{\Phi x}_k+\bm{\Gamma U}_k,\label{eq:basic_mpc_b}\\
		&\bm{X}_k \in \underbrace{\mathbb{X}\times\dots\times\mathbb{X}}_{N-1 \text{ times}}\times\mathbb{X}_T&=: \mathcal{X}_N,\label{eq:basic_mpc_d}\\
		&\bm{U}_k \in \underbrace{\mathbb{U}\times\dots\times\mathbb{U}}_{N \text{ times}}&=:\mathcal{U}_N,\label{eq:basic_mpc_e}
	\end{align}
\end{subequations}
where
\begin{subequations}
	\begin{align}
		&J(\bm{x}_k,\bm{U}_k) := \ell_T(\bm{x}_{N|k}) + \textstyle\sum_{i=0}^{N-1}\ell(\bm{x}_{i|k},\bm{u}_{i|k}),\\
		&\bm{X}_k:=[\bm{x}_{1|k}^\top \ \dots\ \bm{x}_{N|k}^\top]^\top,\label{eq:basic_mpc_f}\\
		&\bm{U}_k:=[\bm{u}_{0|k}^\top \ \dots\ \bm{u}_{N-1|k}^\top]^\top,\label{eq:basic_mpc_g}\\
		&\bm{\Phi} := \begin{bmatrix}
			\bm{A} \\ \vspace{-0.15cm}  
			\bm{A}^2 \\
			\vdots \\
			\bm{A}^N
		\end{bmatrix},\ \bm{\Gamma}:=\begin{bmatrix}
		\bm{B} & \bm{0} & \hspace{-5pt}\dots & \hspace{-5pt}\bm{0}\\ \vspace{-0.15cm}
		\bm{AB} & \bm{B} & \hspace{-5pt}\dots & \hspace{-5pt}\bm{0} \\ 
		\vdots & \vdots & \hspace{-5pt}\ddots   & \hspace{-5pt}\vdots \\
		\bm{A}^{N-1}\bm{B} & \hspace{-5pt}\bm{A}^{N-2}\bm{B} & \hspace{-5pt}\dots &\hspace{-5pt}\bm{B} 
	\end{bmatrix}.
	\end{align}
\end{subequations}
Here, $\ell(\cdot)$, $\ell_T(\cdot)$, $\bm{x}_{i|k}$, $\bm{u}_{i|k}$, and $\mathbb{X}_T$ denote the stage cost, the terminal cost, the predicted state, the predicted input, and the polyhedral terminal set, respectively. The terminal set satisfies ${\mathbb{X}_T:=\{\bm{x}\in\mathbb{R}^n\mid\bm{C}_T\bm{x}\leq\bm{b}_T\}\subseteq\mathbb{X}}$, where ${\bm{C}_T\in\mathbb{R}^{n_T\times n}}$ and ${\bm{b}_T\in\mathbb{R}^{n_T}}$, which is positively invariant with respect to $\mathbb{U}$ for some auxiliary control law $\bm{u} = \bm{Kx}$, i.e., $(\bm{A} + \bm{BK})\mathbb{X}_T\subseteq\mathbb{X}_T$ and $\bm{K}\mathbb{X}_T\subseteq\mathbb{U}$. The $i|k$ subscript is used to denote the $i$-th prediction at time $k$. For example, $\bm{x}_{i|k}$ denotes the prediction of $\bm{x}_{k+i}$ made at time $k$ for $i=1,2,\dots,N$, where $N$ denotes the prediction horizon. Additionally, the stage and terminal cost quantify the performance in terms of the predicted states $\bm{x}_{i|k}$ and inputs $\bm{u}_{i-1|k}$ for $i=1,2,\ldots,N$. 

For the optimization problem \eqref{eq:basic_mpc}, we denote the set of feasible inputs parameterized by the state $\bm{x}_k$ by
\begin{align}
	\mathcal{U}_{f,N}(\bm{x}_k) :=\{ \bm{U}_k \in \mathcal{U}_N \mid \eqref{eq:basic_mpc_b}-\eqref{eq:basic_mpc_e}\}.
\end{align}
Moreover, we denote the set of feasible states by 
\begin{align} \label{eq:feasible}
	\mathbb{X}_f :=\{ \bm{x} \in \mathbb{X} \mid \mathcal{U}_{f,N}(\bm{x}) \neq \emptyset \}.
\end{align}
Finally, note that under suitable assumptions on $\ell(\cdot)$ and $\ell_T(\cdot)$, for all $\bm{x}_k\in \mathbb{X}_f$ a minimizer of \eqref{eq:basic_mpc} exists and we denote by $\bm{U}^\star_{k} := [\bm{u}^{\star\top}_{0|k}\ \dots\ \bm{u}^{\star\top}_{N-1|k}]^\top$ a particular one at time $k\in\mathbb{N}$, i.e., 
\begin{align}\label{eq:argmin}
	\bm{U}^\star_{k} \in \underset{\bm{U}_k \in \mathcal{U}_{f,N}(\bm{x}_k)}{\arg \min} J(\bm{x}_k,\bm{U}_k).
\end{align}
The optimal predicted state sequence corresponding to $\bm{U}^\star_{k}$ is denoted by ${\bm{X}_k^\star = [\bm{x}_{1|k}^{\star\top}\ \dots\ \bm{x}_{N|k}^{\star\top}]^\top = \bm{\Phi x}_k + \bm{\Gamma U}_k^\star}$. 

Using a receding horizon implementation, the MPC problem \eqref{eq:basic_mpc} is turned into a feedback law $K_\text{MPC}:\mathbb{X}_f\rightarrow\mathbb{U}$ by applying the first computed input in $\bm{U}_k^\star$ on the real plant \eqref{eq:system_description}, i.e., $\bm{u}_k := K_\text{MPC}(\bm{x}_k):= \bm{u}^\star_{0|k}$.


\subsection{Forward and backward reachable sets}\label{sec:reachable}
We define the forward reachable set $\overrightarrow{\mathcal{R}}^{k}(\mathbb{X}_0,\mathbb{U})$ for initial state set $\mathbb{X}_0\subseteq \mathbb{R}^n$ as the collection of all states reachable from states in $\mathbb{X}_0$ in $k\in\mathbb{N}$ time-steps subject to the system dynamics \eqref{eq:system_description} and input constraints \eqref{eq:input_constraints}. More precisely, the forward reachable set for \eqref{eq:system_description}, \eqref{eq:input_constraints}, and initial state set $\mathbb{X}_0$ can be computed using the recursion
\begin{align}
	\overrightarrow{\mathcal{R}}^{k+1}(\mathbb{X}_0,\mathbb{U}) := &\{\bm{x}^+\in\mathbb{R}^n\mid \bm{x}^+=\bm{Ax}+\bm{Bu},\\ 
	\text{for some}&\ \bm{x}\in\overrightarrow{\mathcal{R}}^{k}(\mathbb{X}_0,\mathbb{U}),\text{ and } \bm{u}\in\mathbb{U}\},\ k\in\mathbb{N}, \nonumber 
\end{align}
where $\overrightarrow{\mathcal{R}}^{0}(\mathbb{X}_0,\mathbb{U}) := \mathbb{X}_0$. Similarly, the backward reachable set $\overleftarrow{\mathcal{R}}^{k}(\mathbb{X}_T,\mathbb{U})$ for a given terminal set at time $N\in\mathbb{N}$, $\mathbb{X}_T\subseteq\mathbb{X}$, is obtained through a backwards recursion
\begin{align}\label{eq:bwd}
	\overleftarrow{\mathcal{R}}^{k-1}(\mathbb{X}_T,\mathbb{U}) &:= \{\bm{x}\in\mathbb{R}^n\mid\bm{Ax}+\bm{Bu} \in \overleftarrow{\mathcal{R}}^{k}(\mathbb{X}_T,\mathbb{U})\\
	&\text{for some}\ \bm{u}\in\mathbb{U}\},\ k\in\mathbb{N}_{[1,N]},\nonumber
\end{align}
where $\overleftarrow{\mathcal{R}}^{N}(\mathbb{X}_T,\mathbb{U}):= \mathbb{X}_T$.
Note that the definitions for the forward and backward reachable sets \textit{do not include} the state constraints, but do include the input constraints. 

Last, for ease of notation, we define forward and backward reachable tubes as
\begin{subequations}\label{eq:reachable_tubes}
	\begin{align}
		&\overrightarrow{\mathcal{R}}(\mathbb{X}_0,\mathbb{U}) := \overrightarrow{\mathcal{R}}^{1}(\mathbb{X}_0,\mathbb{U}) \times \dots \times \overrightarrow{\mathcal{R}}^{N}(\mathbb{X}_0,\mathbb{U}),\\
		&\overleftarrow{\mathcal{R}}(\mathbb{X}_T,\mathbb{U}) := \overleftarrow{\mathcal{R}}^{1}(\mathbb{X}_T,\mathbb{U}) \times \dots \times \overleftarrow{\mathcal{R}}^{N-1}(\mathbb{X}_T,\mathbb{U})\times \mathbb{R}^n.\label{eq:reachable_tubes_bwd}
	\end{align}
\end{subequations}
Note that, in \eqref{eq:reachable_tubes_bwd} we replace $\overleftarrow{\mathcal{R}}^N(\mathbb{X}_T,\mathbb{U})$ with $\mathbb{R}^n$ to ensure we do not remove terminal constraints with $\mathbb{X}_T$ itself.


\subsection{Cost function level set}
Besides the forward and backward reachable sets, we will also exploit the cost function level set to remove constraints from \eqref{eq:basic_mpc_d}. The key observation is that for state $\bm{x}_k\in\mathbb{X}_f$ and any given input sequence $\tilde{\bm{U}}_k\in\mathcal{U}_{f,N}(\bm{x}_k)$, an upper bound for the MPC cost function can be easily obtained as $J(\bm{x}_k,\bm{U}^\star_k)\leq J(\bm{x}_k,\tilde{\bm{U}}_k)$. More precisely, this upper bound defines a cost function level set
\begin{align}\label{eq:costfun_levelset}\hspace{-5pt}
	\mathcal{J}_u(\bm{x}_k,\tilde{\bm{U}}_k):=&\{\bm{U}\in\mathbb{R}^{Nm} \mid J(\bm{x}_k,\bm{U}) \leq J(\bm{x}_k,\tilde{\bm{U}}_k)\}. 
\end{align}
For the well-known case where $J(\bm{x},\bm{U})$ is a quadratic cost function such that
\begin{align}
	\ell(\bm{x},\bm{u}) = \bm{x}^\top\bm{Q}\bm{x} + \bm{u}^\top\bm{R}\bm{u},\quad \ell_T(\bm{x}) = \bm{x}^\top\bm{P}\bm{x},
\end{align}
with $\bm{Q},\ \bm{R}$, and $\bm{P}$ positive-definite matrices, the resulting level set \eqref{eq:costfun_levelset} is an ellipse in $\mathbb{R}^{Nm}$ as a function of $\bm{x}_k$, $\tilde{\bm{U}}_k$.

Using \eqref{eq:basic_mpc_b}, it is also possible to define a cost function level set for state trajectories in $\mathbb{R}^{Nn}$, 
\begin{align}
\mathcal{J}_x(\bm{x}_k,\tilde{\bm{U}}_k) :=\{\bm{\Phi x}_k+\bm{\Gamma U}\mid \bm{U}\in\mathcal{J}_u(\bm{x}_k,\tilde{\bm{U}}_k)\}.
\end{align}

\section{Constraint removal using reachable sets and cost function level sets}\label{sec:reduction}
Generally, given a convex MPC problem, it is known that there exist exactly at most $Nm$ so-called support constraints whose removal changes the minimizer \cite{Boyd2009}. However, removing all but these state-dependent support constraints is time-consuming. Nevertheless, it is possible to improve the computational complexity of \eqref{eq:basic_mpc} by removing only a subset of the non-support constraints. Clearly, this speeds up the MPC problem without impacting the closed-loop performance, recursive feasibility, and constraint satisfaction. 


We start with the observation that we can remove state constraints that are either, not forward- or backward-reachable, or correspond to state trajectories that do not lower the MPC cost below $J(\bm{x},\tilde{\bm{U}}_k)$. To start exploiting this observation, we define the set
\begin{align}\label{eq:def_H}
	\mathcal{H}(\bm{x}_k,\tilde{\bm{U}}_k) := \overrightarrow{\mathcal{R}}(\bm{x}_k,\mathbb{U}) \cap \overleftarrow{\mathcal{R}}(\mathbb{X}_T,\mathbb{U})\cap\mathcal{J}_x(\bm{x}_k,\tilde{\bm{U}}_k).
\end{align}
\begin{definition}\label{def:red_constraint}
	A constraint $\bm{c}_j\bm{x}\leq b_j$ is called ($\bm{x}_k,\ \mathbb{U},\ \mathbb{X}_T,\ \tilde{\bm{U}}_k$)-\textit{redundant} (or \textit{redundant} for short) at time $i\in\mathbb{N}_{[1,N]}$ for system \eqref{eq:system_description}, if
	\begin{align}\label{eq:red_constraint}
		&[\bm{x}_{1|k}^\top \ \dots\ \bm{x}_{N|k}^\top]^\top \in \mathcal{H}(\bm{x}_k,\tilde{\bm{U}}_k)\implies  \bm{c}_j\bm{x}_{i|k}\leq b_j.
	\end{align}
\end{definition}
Loosely speaking, the constraint $\bm{c}_j\bm{x}_{i|k}\leq b_j$, where $\bm{C}:=[\bm{c}_1^\top,\dots,\bm{c}_{n_x}^\top]^\top$ and $\bm{b}:=[b_1,\dots,b_{n_x}]^\top$, is redundant at time $i$, if system dynamics, input constraints, terminal set, and MPC cost function already guarantee its satisfaction. Clearly, if $\bm{c}_j\bm{x}_{i|k}\leq b_j$ is redundant at time $i$, it can be removed from the MPC problem without changing the minimizer. 
\begin{definition}
	A set ${\mathcal{X}_N^\text{red}:=\mathbb{X}^\text{red}_1\times\dots\times\mathbb{X}^\text{red}_N}$ is called a \textit{reduced constraint set} for $\mathcal{X}_N$, if there are index sets ${\mathbb{A}_i\subseteq\ \mathbb{N}_{[1,n_x]}}$, for $i=1,2,\dots,N-1$, such that
	\begin{subequations}\label{eq:def_red_constraint_set}
	\begin{align}
		&\mathbb{X}^\text{red}_i=\{\bm{x}\in\mathbb{R}^n\mid \bm{c}_j\bm{x}\leq b_j,\ \text{for }j\in\mathbb{A}_i\}.
	\end{align}
For $i=N$, we define index sets $\mathbb{A}_N\subseteq\mathbb{N}_{[1,n_T]}$ such that
	\begin{align}
	&\mathbb{X}^\text{red}_N=\{\bm{x}\in\mathbb{R}^n\mid \bm{c}_{T,j}\bm{x}\leq b_{T,j},\ \text{for }j\in\mathbb{A}_N\},
	\end{align}
\end{subequations}
where $\bm{C}_T:=[\bm{c}_{T,1}^\top,\dots,\bm{c}_{T,n_T}^\top]^\top$ and $\bm{b}_T:=[b_{T,1},\dots,b_{T,n_T}]^\top$.
\end{definition}
From the above reasoning the next result follows.
\begin{proposition}\label{prop:1}
	 Consider a reduced constraint set $\mathcal{X}_N^\text{red}$ for $\mathcal{X}_N$. If for initial state $\bm{x}_k\in\mathbb{X}_f$ and $\tilde{\bm{U}}_k\in\mathcal{U}_{f,N}(\bm{x}_k)$ it holds that
	\begin{align}\label{eq:prop1}
		\mathcal{H}(\bm{x}_k,\tilde{\bm{U}}_k) \cap \mathcal{X}_N = 
		\mathcal{H}(\bm{x}_k,\tilde{\bm{U}}_k) \cap \mathcal{X}^\text{red}_N,
	\end{align}
	then the MPC problem, where \eqref{eq:basic_mpc_d} is replaced by $\bm{X}_k\in\mathcal{X}_N^\text{red}$, has the same minimizer as the original MPC problem, i.e.,
	\begin{align}\label{eq:argmin_same}
		\underset{\bm{U}_k \in \mathcal{U}_{f,N}(\bm{x}_k)}{\arg \min} J(\bm{x}_k,\bm{U}_k)
		= \underset{\substack{ \eqref{eq:basic_mpc_b},\ \bm{U}_k\in\mathcal{U}_N\, \bm{X}_k\in\mathcal{X}_N^\text{red}}}{\arg \min} \hspace{-0.5cm} J(\bm{x}_k,\bm{U}_k).
	\end{align}
\end{proposition}


\section{Numerically efficient constraint reduction}\label{sec:removal}
In this section, we provide numerically efficient procedures to remove state constraints based on the observations in Section~\ref{sec:reduction}. While Section~\ref{sec:reduction} describes conditions that can be used to remove redundant constraints, we need efficient and fast techniques to carry out the corresponding calculations online. To this end, we first show that \textit{outer} approximations of the reachable sets and the cost function level set can be exploited in our ca-MPC framework. Second, we provide numerically efficient methods for computing such outer approximations of reachable sets. Finally, we present a method to efficiently construct suitable reduced constraints sets $\mathcal{X}_N^\text{red}$ based on these computational methods.
\subsection{Outer approximations in constraint removal}
In this section, we show that using outer approximations of $\mathcal{H}(\bm{x}_k,\tilde{\bm{U}}_k)$ can be exploited in our framework.
\begin{corollary}
	Consider an outer approximation of $\mathcal{H}(\bm{x}_k,\tilde{\bm{U}}_k)$ in the sense that ${\mathcal{H}(\bm{x}_k,\tilde{\bm{U}}_k)\subseteq\mathcal{H}_+(\bm{x}_k,\tilde{\bm{U}}_k)}$. If 
	\begin{align}\label{eq:h_cap}
		\mathcal{H}_+(\bm{x}_k,\tilde{\bm{U}}_k) \cap \mathcal{X}_N &= \mathcal{H}_+(\bm{x}_k,\tilde{\bm{U}}_k) \cap\mathcal{X}_N^\text{red}
	\end{align}
	holds for a reduced constraint set $\mathcal{X}_N^\text{red}$ of $\mathcal{X}_N$, then \eqref{eq:prop1} holds and thus \eqref{eq:argmin_same}.
\end{corollary}
Clearly, $\mathcal{X}_N^\text{red}$ obtained by removing constraints based on the outer approximation $\mathcal{H}_+(\bm{x}_k,\tilde{\bm{U}}_k)$ has an equal or larger number of constraints than a reduced constraint set that is computed without outer approximations. However, outer approximations can be exploited in our framework by providing a beneficial trade-off between the computational effort of removing redundant constraints and the associated speedup of the MPC problem.

\subsection{Reachable set outer approximations}
Computing reachable sets is known to be a computationally intensive task. However, we will exploit an important feature of reachable sets for LTI systems that will make these computations tractable. For ease of exposition, we limit ourselves to the forward reachable set in the explanation below, but the same properties apply to the backward reachable set, mutatis mutandis.

A key feature of reachable sets for LTI systems is the decomposition $\overrightarrow{\mathcal{R}}(\bm{x},\mathbb{U}) =\bm{\Phi x} + \overrightarrow{\mathcal{R}}(\bm{0},\mathbb{U})$. Here, $\bm{\Phi x}$ is the free state response with initial condition $\bm{x}$ and zero input sequence, while $\overrightarrow{\mathcal{R}}(\bm{0},\mathbb{U})$ is the forward reachable tube starting at the origin. Crucially, $\overrightarrow{\mathcal{R}}(\bm{0},\mathbb{U})$ does \textit{not} depend on the current state and can therefore be computed offline, thereby lowering the online computational burden drastically. Even though the forward reachable tube can be computed offline, it is not necessarily easy and the complexity of the set itself can be intractable in some cases. In such cases, the use of outer approximations of reachable sets can be useful and render the computations tractable while simultaneously lowering the complexity of the set itself. The reduced complexity of the sets themselves will be extensively used to adaptively remove redundant constraints in our ca-MPC framework detailed in Section~\ref{sec:selection}. 

In this paper, we limit ourselves to bounding boxes and outer ellipsoidal approximations of the reachable sets, although the main ideas also apply to other shapes. A bounding box approximates the forward reachable tube such that
\begin{subequations}
	\begin{align}
		&\mathcal{B}(\bm{P}_1,\bm{q}_1,\bm{l}_1)\times \dots \times \mathcal{B}(\bm{P}_N,\bm{q}_N,\bm{l}_N) \supseteq \overrightarrow{\mathcal{R}}(\bm{0},\mathbb{U}),\\
		&\mathcal{B}(\bm{P},\bm{q},\bm{l}):=\{\bm{x}\in\mathbb{R}^n\mid -\bm{l}\leq\bm{P}(\bm{x} - \bm{q})\leq\bm{l}\}.\label{eq:def_ball}
	\end{align}
\end{subequations}
Here, $\bm{P}_1,\dots,\bm{P}_N$ are invertible matrices and $\bm{l}_1,\dots,\bm{l}_N$ non-negative vectors of appropriate dimension. Similarly, an outer ellipsoidal approximation satisfies
\begin{subequations}
	\begin{align}
		&\mathcal{E}(\bm{L}_1,\bm{q}_1)\times \dots \times \mathcal{E}(\bm{L}_N,\bm{q}_N) \supseteq \overrightarrow{\mathcal{R}}(\bm{0},\mathbb{U}),\\
		&\mathcal{E}(\bm{L},\bm{q}):=\{\bm{x}\in\mathbb{R}^n\mid \|\bm{L}^\top(\bm{x}-\bm{q})\|_2 \leq 1\}.
	\end{align}
\end{subequations}
Here, $\bm{L}_1,\dots,\bm{L}_N$ are positive-definite matrices and $\bm{q}_1,\dots,\bm{q}_N$ vectors of appropriate dimension. These outer approximations allow for simple set descriptions that can be used to remove redundant constraints efficiently as we will show. In particular, recent developments in ellipsoidal methods allow for numerically efficient reachable set computations, even for large-scale systems \cite{Halder2018}.

\subsection{Numerically efficient constraint removal}\label{sec:selection}
In this section, we use the previously introduced outer approximations to derive efficient methods to identify redundant state constraints. Recall, that redundant constraints can be identified by verifying if the reduced constraint set satisfies \eqref{eq:h_cap}. One approach to identify and remove a redundant constraint is by solving an optimization problem for each inequality constraint \cite{Paulraj2010}. However, these optimization problems are computationally expensive, therefore, we limit ourselves to numerically efficient, albeit more conservative tests. More specifically, we remove an inequality constraint given by $\bm{cX}\leq b$, when 
\begin{align}\label{eq:conservative_test}
	\{\bm{X}\in\mathbb{R}^{Nn}\mid\bm{cX}\leq b\}\cap \mathcal{H}(\bm{x}_k,\tilde{\bm{U}}_k) = \mathcal{H}(\bm{x}_k,\tilde{\bm{U}}_k)
\end{align}
Checking \eqref{eq:conservative_test} is equivalent to verifing
\begin{align}\label{eq:conservative_test2}
	\bm{X}\in\mathcal{H}(\bm{x}_k,\tilde{\bm{U}}_k)\implies \bm{cX}\leq b.
\end{align}
Next, we elaborate on \eqref{eq:conservative_test2} for box-shaped and ellipsoidal sets, respectively. At the end of this section, we also discuss how a finite intersections and products of boxes and ellipses can be handled efficiently. 

In the box case, we would like to test if 
\begin{align}\label{eq:abstract_box_test}
	\bm{X}\in\mathcal{B}(\bm{P},\bm{q},\bm{l}) \implies \bm{cX}\leq b,
\end{align}
where $\mathcal{B}(\bm{P},\bm{q},\bm{l})$ is an $Nn$-dimensional box-shaped set. In case \eqref{eq:abstract_box_test} holds, we set the test outcome of $\mathcal{M}_B(\bm{c},b,\mathcal{B}(\bm{P},\bm{q},\bm{l}))$ to true (the constraint can be removed), otherwise we set it to false. Elementary calculations show that \eqref{eq:abstract_box_test} is true if and only if
\begin{align}
	\bm{c}\bm{P}^{-1}(\text{sign}(\bm{P}^{-\top}\bm{c}^\top)\circ\bm{l})\leq| b-\bm{cq}|,
\end{align}
where sign($\cdot$) denotes the element-wise sign and $\circ$ the element wise product. Similar, for ellipses we have
\begin{align}\label{eq:abstract_ellipse_test}
	\bm{X}\in\mathcal{E}(\bm{L},\bm{q}) \implies \bm{cX}\leq b,
\end{align}
and thus the constraint can be removed if \eqref{eq:abstract_ellipse_test} holds. Again, elementary calculations show that \eqref{eq:abstract_ellipse_test} holds if and only if
\begin{align}\label{eq:ellipse_test}
	\|\bm{c}\bm{L}^{-\top}\|_2 \leq  |b - \bm{c}\bm{q}|.
\end{align}
Hence, we define $\mathcal{M}_E(\bm{c},b,\mathcal{E}(\bm{L},\bm{q})) \iff \eqref{eq:ellipse_test}$.

Based on the definition of $\mathcal{H}(\bm{x}_k,\tilde{\bm{U}}_k)$ \eqref{eq:def_H}, it is convenient to extend our constraint removal framework to intersections of boxes and ellipses,
\begin{align}\label{eq:h_param}\allowdisplaybreaks
	\mathcal{H}(\bm{x}_k,\tilde{\bm{U}}_k) = \textstyle\bigcap_{p\in\mathbb{N}_{[1,M]}}\mathcal{H}(p,\bm{x}_k,\tilde{\bm{U}}_k),
\end{align}
where $\mathcal{H}(p,\bm{x}_k,\tilde{\bm{U}}_k)$ is either a box or an ellipse for $p\in\mathbb{N}_{[1,M]}$. Observe that when a constraint $\bm{cX}\leq b$ is redundant for some $p\in\mathbb{N}_{[1,M]}$, i.e., $\bm{X}\in\mathcal{H}(p,\bm{x}_k,\tilde{\bm{U}}_k) \implies \bm{c}\bm{X}\leq b$, then \eqref{eq:conservative_test2} also holds. Hence, the constraint can be removed.

Crucially, \eqref{eq:h_param} enables the removal of redundant constraints using simpler and faster, sequential redundancy checks based on $\mathcal{H}(p,\bm{x}_k,\tilde{\bm{U}}_k)$. Additionally, when $\mathcal{H}(\bm{x}_k,\tilde{\bm{U}}_k)$ is given by
\begin{align}\label{eq:h_param2}
	\mathcal{H}(\bm{x}_k,\tilde{\bm{U}}_k) = \mathcal{H}^1(\bm{x}_k,\tilde{\bm{U}}_k)\times\dots\times\mathcal{H}^N(\bm{x}_k,\tilde{\bm{U}}_k),
\end{align} 
as is for the reachable tubes \eqref{eq:reachable_tubes}, then it is also possible to process $\mathcal{H}^i(\bm{x}_k,\tilde{\bm{U}}_k)$ and corresponding constraint set $\mathbb{X}_i^\text{red}$ sequentially for each time $i=1,2,\dots,N$.


\section{Case study: Hyperthermia Treatment}\label{sec:casestudy}
In this section, we demonstrate our proposed ca-MPC framework for numerically efficient constraint removal with a hyperthermia cancer treatment case study. A hyperthermia treatment involves heating a tumor thereby inducing a local fever. As a result, the tumor is sensitized to radio and chemo therapies, making these therapies more effective \cite{Deenen2020,Hendrikx2018,VanderZee2002,Hensley2015}. Critically, tissue temperature constraints ensure the safety and comfort of the patient \cite{Yarmolenko2011}. The temperature constraint is defined on the \textit{continuous} spatial domain and therefore results in many state constraints after spatial discretization, which is a typical scenario for MPC-based PDE control.

The remainder of this section is structured as follows. First, the system dynamics and MPC problem are introduced. Second, the ca-MPC relevant sets are computed. Last, we compare the results obtained using ca-MPC and the original non-reduced MPC setup.
\subsection{System description}
We model the system dynamics in the hyperthermia case study using the 1D heat equation with distributed control, convective boundary conditions, and additional convection on the spatial domain to model the body's thermoregulatory response. This leads for the normalized spatial coordinates $r\in[0,1]$ and time $t\in\mathbb{R}_{\geq0}$ to
\begin{subequations}\label{eq:pde}
	\begin{align}
		&\dot{T}(r,t) = \alpha \nabla^2T(r,t) - \beta T(r,t) + Q(r,t),\\
		&Q(r,t) = B_1(r)u_1(t) + B_2(r)u_2(t)
	\end{align}
\end{subequations}
with boundary conditions
\begin{align}\label{eq:bc}
	\nabla T(0,t) = \gamma T(0,t),\quad 	\nabla T(1,t) = -\gamma T(1,t).
\end{align}
The parameters are given by $\alpha = 2.5\cdot10^{-4}$, $\beta = 10^{-2}$, and $\gamma = 2.5\cdot10^{-3}$. Note that the 1D setup enables the comparison with a non-reduced MPC setup, which would become intractable for 2D and 3D cases. 

\begin{figure}[!ht]
	\centering
	\vspace{-0.2cm}
	\includegraphics[width=8.85cm]{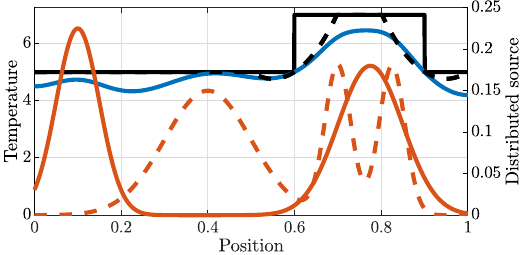}
	\vspace{-0.75cm}
	\caption[]{Temperature constraint $T_\text{max}(r)$ (\tikz{\draw[color=white] (0pt,0pt) -- (15pt,0pt);\draw[ultra thick] (0pt,2.5pt) -- (15pt,2.5pt);}), terminal constraint $T_\text{terminal}(r)$ (\tikz{\draw[color=white] (0pt,0pt) -- (15pt,0pt);\draw[ultra thick, dashed] (0pt,2.5pt) -- (15pt,2.5pt);}), the state reference $\bm{x}_\text{r}$ (\tikz{\definecolor{blue}{HTML}{0088cd};\draw[color=white] (0pt,0pt) -- (15pt,0pt);\draw[ultra thick, color=blue] (0pt,2.5pt) -- (15pt,2.5pt);}), and the distributed source $B_1(r)$ (\tikz{\definecolor{red}{HTML}{d95319};\draw[color=white] (0pt,0pt) -- (15pt,0pt);\draw[ultra thick, color=red] (0pt,2.5pt) -- (15pt,2.5pt);}), and $B_2(r)$ (\tikz{\definecolor{red}{HTML}{d95319};\draw[color=white] (0pt,0pt) -- (15pt,0pt);\draw[ultra thick, color=red,dashed] (0pt,2.5pt) -- (15pt,2.5pt);}), for $r\in[0,1]$.}
	\label{fig:constraints}\end{figure}

To ensure patient safety and efficacy of the treatment, we constrain the healthy tissue temperature increase to five degrees with respect to the body temperature and seven degrees for the tumor on $r\in[0.6,0.9]$. This constraint is illustrated in Fig.~\ref{fig:constraints} by the function $T_\text{max}(r)$.

Heating is applied to the patient using RF-applicator \cite{Hendrikx2018}, which is modeled through the actuator profiles $B_1(r)$ and $B_2(r)$, see Fig.~\ref{fig:constraints}. Observe that the heat deposition is not limited to the tumor region, this is also typically observed in real-RF-based hyperthermia treatments. Healthy tissue regions with high power deposition are typically referred to as hot spots. These hot spots also motivate the use of control strategies that explicitly include state constraints to ensure patient safety. To ensure that the actuator applies no negative heat load and respects power limits, the input constraints are given by $	0\leq u_1 \leq 1$ and $0\leq u_2\leq 1$.

Next, we spatially and temporally discretize \eqref{eq:pde} and \eqref{eq:bc}, using a central-difference scheme and zero-order hold, respectively, to obtain a model as in \eqref{eq:system_description}. Here, each element of $\bm{x}_k\in\mathbb{R}^n$ denotes a nodal temperature defined on an equidistant grid of $n$ points, i.e., ${\{0,\frac{1}{n-1},\frac{2}{n-1},\dots,1\}}$. Note that the case study will use a varying number of discretization points $n$ to demonstrate our ca-MPC framework.

In order to heat the tumor, we use a quadratic cost function to track the reference $\bm{x}_r$ illustrated in Fig.~\ref{fig:constraints}. The reference $\bm{x}_\text{r}$ and $\bm{u}_\text{r}$ are obtained through an offline optimization procedure to maximize the tumor temperature. The nominal MPC setup is now given as
\begin{subequations}\label{eq:hypertermia}
	\begin{align}
		\underset{\bm{U}_k}{\text{minimize}}\quad&\textstyle\sum_{i=1}^{10}\|\bm{x}_{i|k}-\bm{x}_\text{r}\|^2_2 + \|\bm{u}_{i-1|k}-\bm{u}_\text{r}\|^2_2,\label{eq:hypertermia_a}\\
		\hspace{-200pt}\text{subject to }\quad&\bm{X}_k = \bm{\Phi x}_k+\bm{\Gamma U}_k,\label{eq:hypertermia_b}\\
		&\bm{x}_{i|k} \leq \bm{T}_\text{max},\quad i = 1,\dots,N-1,\label{eq:hypertermia_d}\\
		&\bm{x}_{N|k} \leq \bm{T}_\text{terminal},\label{eq:hypertermia_e}\\
		&\bm{0}\leq \bm{u}_{i|k} \leq \bm{1},\quad i = 0,\dots,N-1.\label{eq:hypertermia_f}
	\end{align}
\end{subequations}
Here, $\bm{T}_\text{max}\in\mathbb{R}^n$ denotes the spatially discrete temperature constraint, see Fig.~\ref{fig:constraints}. The terminal set constraint \eqref{eq:hypertermia_e} maximizes $\|\bm{T}_\text{terminal}\|_1$ while satisfying $(\bm{A}-\bm{I})\bm{T}_\text{terminal} \leq \bm{0}$ and $\bm{T}_\text{terminal} \leq \bm{T}_\text{max}$ to ensure positive invariance with auxiliary input $\bm{u} =\bm{0}$. For technical details we refer the interested reader to \cite{Bitsoris1988}. 
\scalebox{0}{
	\begin{tikzpicture}
		\begin{axis}[hide axis]
			\addplot [
			color=black,
			solid,
			line width=2pt,
			forget plot
			]
			(0,0);\label{leg:black}
			\addplot [
			color=black,
			dashed,
			line width=2pt,
			forget plot
			]
			(0,0);\label{leg:blackDash}
			\addplot [
			color=blue,
			solid,
			line width=2pt,
			forget plot
			]
			(0,0);\label{leg:blue}
			\addplot [
			color=red,
			solid,
			line width=2pt,
			forget plot
			]
			(0,0);\label{leg:red}
			\addplot [
			color=red,
			dashed,
			line width=2pt,
			forget plot
			]
			(0,0);\label{leg:redDashed}
		\end{axis}
	\end{tikzpicture}%
}
	\vspace{-0.7cm}
	
\subsection{ca-MPC setup}
We approximate the forward and backward reachable sets using axis-aligned boxes and exploit the observation that our model \eqref{eq:pde} is a positive system. As a result, the forward reachable sets are outer-approximated by
\begin{align}\label{eq:hypertermia_fwd}
	\overrightarrow{\mathcal{R}}^{i}(\bm{0},\mathbb{U}) := \{\bm{x}\in\mathbb{R}^n\mid \bm{0}\leq\bm{x}\leq \textstyle\sum_{k=0}^{i-1}\bm{A}^{k} \bm{B}\bm{1}\}.
\end{align}
Here, $\bm{1}\in\mathbb{R}^2$ denotes a column of ones. Moreover, note that \eqref{eq:hypertermia_fwd} can be trivially transformed to \eqref{eq:def_ball}. Conceptually, \eqref{eq:hypertermia_fwd} denotes a set based on the maximum temperature that can be achieved with both inputs. 
Next, the backward reachable set is approximated by
\begin{align}\label{eq:hypertermia_bwd}
	&\overleftarrow{\mathcal{R}}^{i}(\mathbb{X}_T,\bm{0}) := \{\bm{x}\in\mathbb{R}^n\mid 0\leq x_j \leq T_{j,\text{terminal}} + \delta_{i,j},\\
	&\qquad\qquad\qquad\qquad\qquad\qquad
	 \text{for } j=1,\dots,n\},\nonumber
\end{align}
where $\delta_{i,j} := \underset{\bm{A}^{N-i}\bm{e}_j\delta \leq (\bm{I} - \bm{A}^{N-i})\bm{T}_\text{terminal}}{\max\quad \delta}$ and $\bm{e}_j$ denotes the $j$-th standard basis vector. Conceptually, \eqref{eq:hypertermia_bwd} is the union of maximum single-state perturbations that satisfy the terminal set ater $N-i$ time-steps with zero input, i.e., $\bm{u} = \bm{0}$. Again, \eqref{eq:hypertermia_bwd} can be trivially transformed to \eqref{eq:def_ball}.

To obtain the cost function level set, we substitute \eqref{eq:basic_mpc_b} into the MPC cost function \eqref{eq:hypertermia_a}, which yields
\begin{align}
	&J(\bm{x}_k,\bm{U}_k) = \frac{1}{2}\bm{U}_k^\top\bm{GG}^\top \bm{U}_k + \bm{f}^\top(\bm{x}_k)\bm{U}_k + c(\bm{x}_k),
\end{align}
where $\bm{G}$, $\bm{f}(\bm{x}_k)$, and $c(\bm{x}_k)$ denote the respective matrix, vector, and scalar obtained by rewriting \eqref{eq:hypertermia}. Given an initial state $\bm{x}_k\in\mathbb{X}_f$, and input sequence $\tilde{\bm{U}}_k\in\mathcal{U}_{f,N}(\bm{x}_k)$, the cost function level set is given by the ellipse ${\mathcal{J}_u(\bm{x}_k,\tilde{\bm{U}}_k) := \mathcal{E}(\bm{L}_k,\bm{q}_k)}$ with $\bm{L}_k := \bm{G}\|\bm{G}^\top(\tilde{\bm{U}}_k-\bm{q}_k)\|_2^{-1}$ and $\bm{q}_k:= -(\bm{GG}^\top)^{-1}\bm{f}(\bm{x}_k)$.

Our framework allows for any feasible input sequence ${\tilde{\bm{U}}_k\in\mathcal{U}_{f,N}(\bm{x}_k)}$, but a natural choice is a shifted version of the previous optimal sequence and appended by the auxiliary control law, ${\tilde{\bm{U}}_k = [\bm{u}^{\star\top}_{1|k-1}\ \dots\ \bm{u}^{\star\top}_{N-1|k-1}\ \bm{0}^\top]^\top}$. Crucially, note that $\tilde{\bm{U}}_k$ ensures feasibility as the terminal set was chosen to be positively invariant for $\bm{u}=\bm{0}$.

For efficiency reasons, we sequentially remove state constraints for each time step and the forward-, backward-, and cost function level set, as mentioned in \eqref{eq:h_param} and \eqref{eq:h_param2}. More specifically, we compute each reduced constraint set $\mathbb{X}^\text{red}_i$ for $i=1,\dots,N-1$ using
\begin{subequations}\label{eq:real_reduction}
	\begin{align}
		\mathbb{A}_i &= \{j\in\mathbb{N}_{[1,n_x]}\mid \neg\big[\mathcal{M}_B(\bm{c}_j, b_j,\overrightarrow{\mathcal{R}}^i) \lor\\ 
		&\mathcal{M}_B(\bm{c}_j, b_j,\overleftarrow{\mathcal{R}}^i) \lor \mathcal{M}_E(\hat{\bm{c}}_{j,i}, \hat{b}_{j,i},\mathcal{J}_u(\bm{x}_k,\tilde{\bm{U}}_k))\big]\},	\label{eq:input_ellipse}
	\end{align}
and for $i=N$,
	\begin{align}
		\mathbb{A}_N = \{&j\in\mathbb{N}_{[1,n_T]}\mid	\neg\big[\mathcal{M}_B(\bm{c}_{T,j}, b_{T,j},\overrightarrow{\mathcal{R}}^N) \lor\\ 
		&\mathcal{M}_E(\hat{\bm{c}}_{T,j}, \hat{b}_{T,j},\mathcal{J}_u(\bm{x}_k,\tilde{\bm{U}}_k))\big]\}.\label{eq:input_ellipse2}	
	\end{align}
\end{subequations}
Here $\bm{c}_j$, $b_j$, $\bm{c}_{T,j}$, $b_{T,j}$ denote the state constraints $\bm{x}\leq \bm{T}_\text{max}$ and $\bm{x}\leq \bm{T}_\text{terminal}$ written as $\bm{Cx}\leq\bm{b}$ and $\bm{C}_T\bm{x}\leq\bm{b}_T$, respectively. The operator $\lor$, denotes the logical OR, and $\neg$ the logical NOT. For \eqref{eq:input_ellipse} and \eqref{eq:input_ellipse2}, $\hat{\bm{c}}_{j,i}$ and $\hat{b}_{j,i}$ denote the state constraint $\bm{c}_j\bm{x}_{i|k}\leq b_j$ at time $i$ as $\hat{\bm{c}}_{j,i}\bm{U}_k\leq \hat{b}_{j,i}$ using \eqref{eq:basic_mpc_b}. This allows for $\mathcal{M}_E(\cdot)$ to be evaluated using the $Nm$-dimensional $\bm{U}_k$-space in contrast to the $Nn$-dimensional $\bm{X}_k$-space, which is typically faster when $m < n$. 
 

\subsection{Results}
\begin{figure}[!ht]
    \vspace{-0.1cm}
	\centering
	\includegraphics[width=8.85cm]{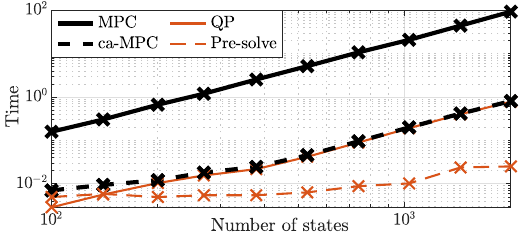}
	\vspace{-0.6cm}
	\caption{The maximum computation time of our constraint-adaptive MPC method and a standard MPC for increasingly fine spatial discretizations. For our framework the solver and pre-solve time are also shown.}
	\label{fig:cpu_time}
\end{figure}

To illustrate the obtained computational improvements we compare the maximum computation time of our proposed ca-MPC framework with respect to a standard MPC setup, with an increasing number of states $n\in[100,2000]$. Additionally, for our proposed method, we differentiate between the computation time of the pre-solve and quadratic program to provide insight in the overhead of our method. In Fig.~\ref{fig:cpu_time}, the resulting computation time is shown. Clearly, our framework can achieve speedups of two-orders of magnitude. Moreover, the overhead from the pre-solve step scales well with an increasing number of constraints.

\begin{figure}[!ht]
	\vspace{-0.4cm}
	\centering
	\includegraphics[width=8.85cm]{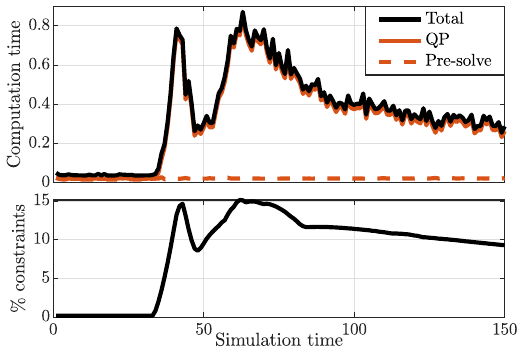}
	\vspace{-0.6cm}
	\caption{Computation time for ca-MPC and the percentage of constraints that are included in the MPC problem for $n=2000$. The computation time correlates well with the number of constraints. Moreover, the pre-solve overhead remains approximately constant throughout the simulation.}
	\label{fig:campc_time}
\end{figure}

Finally, in Fig.~\ref{fig:campc_time} the computation time as well as the percentage of total constraints are shown for a single ca-MPC simulation with $n=2000$. Here, the adaptive nature of the method is clearly seen as the number of included constraints varies during the simulation. During the first 30 seconds, barely any constraints are included as the state cannot reach the constraints within the prediction horizon. Hereafter, the system approaches the temperature constraints, which results in the need to include more constraints as seen in Fig.~\ref{fig:campc_time}. Healthy tissue temperatures approaching the temperature constraints are also typical for real hyperthermia treatments due to the presence of hot spots. After approximately 60 seconds, the cost function level set starts to shrink as the system approaches the steady-state reference. As result, more constraints can be removed (due to the cost level set) even though the constraints are approached even closer.
\vspace{-0.3cm}

\section{Conclusion and Outlook}\label{sec:conclusion}
In this paper, we presented a novel \textit{constraint-adaptive} MPC scheme that can enable real-time feasibility for systems with many state constraints. Crucially, our framework guarantees the same closed-loop performance, recursive feasibility and constraints satisfaction properties as the non-reduced MPC scheme. The presented method exploits fast outer approximation calculations of reachable sets as-well as the cost function level set, to efficiently remove redundant constraints in an online setting. Moreover, the majority of the computational complexity is offline by exploiting key properties of reachable sets for LTI systems. Hence, we believe the method is applicable to a large variety of systems with many state constraints. A hyperthermia cancer treatment case study demonstrated the benefits of ca-MPC showing up to two orders of magnitude reduction in computation time with respect to a standard MPC setup without loss of closed-loop performance.

This paper focused on computationally cheap methods in order to obtain low computational overhead for the constraint removal pre-solve step. Clearly, less conservative checks can be performed to remove more redundant constraints. For example, state constraints that are not removed are currently not used to identify more redundant constraints. Furthermore, specific solver interactions with, e.g., interior-point and active-set solvers are of interest. These are topics for future work next to applying our framework to a full (3D) application in simulation and experiments of a real hyperthermia treatment.

\bibliographystyle{IEEEtran}
\bibliography{literature}
\end{document}